# Time-resolved electron beam diagnostics with sub-femtosecond resolution

**Guanglei Wang[1], Meng Zhang[2], Wei Zhang[2], Haixiao Deng[2*], Xueming Yang[1]**

[1]State Key Laboratory of Molecular Reaction Dynamics, Dalian Institute of Chemical Physics, Chinese Academy of Sciences, Dalian 116023, P. R. China
[2]Shanghai Institute of Applied Physics, Chinese Academy of Sciences, Shanghai, 201800, P. R. China
[*]denghaixiao@sinap.ac.cn

## ABSTRACT

In modern high-gain free-electron lasers, ultra-fast photon pulses designed for studying chemical, atomic and biological systems are generated from a serial of behaviors of high-brightness electron beam at the time-scale ranging from several hundred femtoseconds to sub-femtosecond. Currently, radiofrequency transverse deflectors are widely used to provide reliable, single-shot electron beam phase space diagnostics, with a temporal resolution of femtosecond. Here, we show that the time resolution limitations caused by the intrinsic beam size in transverse deflectors, can be compensated with specific transverse-to-longitudinal coupling elements. For the purpose, an undulator with transverse gradient field is introduced before the transverse deflector. With this technique, a resolution of less than 1fs root mean square has been theoretically demonstrated for measuring the longitudinal profile and/or the micro-bunching of the electron bunch.

## Introduction

Ultra-short pulse duration in the femtosecond range is one of the essential characteristics of the extreme ultraviolet (EUV) to X-ray free-electron lasers (FELs), which may enable the time-resolved study of electronic dynamics[1,2] and allow novel approaches in structure determination of bio molecules[3]. An FEL facility mainly consists of an electron beam accelerator in which the bunch length is typically compressed from several picoseconds to femtosecond order[4-9], and an undulator based photon amplifier where the electron beam is micro-bunched at the time-scale of the radiated photon wavelength, i.e., far less than 1 femtosecond[10,11]. It turns out that the accurate knowledge of the electron beam longitudinal distribution is very crucial for performance optimization and reliable operation of an FEL facility, and understanding the physics behind an FEL setup. However, great challenges remain in diagnosing such electron bunches with femtosecond or sub-femtosecond temporal resolution.

Several techniques have been developed over the past decades to characterize the length and temporal profile of short electron bunches (see a recent review in Ref. [12]). One of them is the electro-optical sampling[13,14] where the polarization of a laser propagating in a crystal changes with the presence of the Coulomb field carried by the electron beam. By measuring the field induced birefringence results in the laser, the bunch temporal profile can be retrieved. This is a single-shot method with a best resolution reported to date of about 50 fs[15]. Multi-shot methods such as those based on spectral analysis techniques[16,17] of the coherent terahertz radiation emitted from the electron bunch have also been explored, which typically yield the pulse duration but not the temporal distribution.

A radiofrequency (RF) transverse deflector provides a reliable, single-shot measurement of the electron beam longitudinal phase space with femtosecond resolution[18,19]. The operating principle of the deflector is similar to that of a streak camera. The time variation of RF field deflects the electron bunch from head to tail by different amounts, and thus transfers the longitudinal bunch shape into the transverse plane, which can be measured with a suitable screen and camera downstream. Since the method effectively converts one of the profile monitor's transverse dimensions into time, it allows the measurement of time-resolved charge distribution or other parameters of the electron beam. The transverse deflector is now an essential diagnostic for the commissioning and operation of worldwide FEL facilities. At the LCLS, such a deflector was recently placed downstream from the undulators, which allows continuously monitor the x-ray pulse duration without interfering with FEL operation[20]. Generally speaking, the resolution of a deflector can be improved by increasing the deflection voltage and the frequency of the deflection field. For instance, by utilizing an X-band deflector with deflecting voltage of 45 MV, a time resolution of about 1fs was recently achieved at LCLS[20]. In order to obtain an even higher resolution, an RF energy doubler is under preparation for the LCLS X-band deflector[21]. However, the time resolution of the deflector is still fundamentally limited by the intrinsic transverse emittance of the electron beam. Therefore, in Ref. [22], a sophisticated chicane is suggested to overcome the limitation arising from the beam energy spread and transverse emittance.

Here, we report a simple and straightforward modification to the deflector technique, in order to compensate the temporal resolution limitation caused by the intrinsic beam size. In detail, an additional undulator with transverse gradient magnetic field, known as TGU[23, 24] in short is utilized to introduce transverse-to-longitudinal coupling in the electron beam phase space[25, 26] before the deflector. Using this technique, it is found that a sub-femtosecond temporal resolution can be achieved theoretically with a reduced deflecting voltage or frequency. The proposed technique provide an ideal tool for measuring the longitudinal profile and/or the micro-bunching structures of the electron bunch, and thus studying the FEL lasing evolution and performance optimization. Meanwhile, the tunable gap TGU serves as an infrared afterburner. The 'free' infrared radiation from the same electron bunch enables the novel pump-probe experiments combining coherent infrared pulses with the FEL radiation[27-29], which is out of the scope of this article.

## Results

**TGU enhanced transverse deflector.** When an electron beam passes through the transverse deflector at a zero-crossing phase, a transverse angular kick (here we assume the kick is in horizontal direction and the bunch length is much shorter than the RF field wavelength) that varies linearly with the longitudinal position will be imprinted on the beam. If the initial state of an electron is described as $[x_0, x'_0, y_0, y'_0, z_0, \delta_0]^T$, where $x_0, y_0$ is the transverse position, $x'_0, y'_0$ is the transverse divergence, $z_0$ and $\delta_0$ are relative longitudinal position and energy deviation with respect to the reference particle, after the deflector and some phase advance, the horizontal position of the electron on the diagnostics screen can be written as:

$$x = x_0 + x'_0 R_{12} + z_0 u R_{12}. \tag{1}$$

Here, $u = 2\pi eV/\lambda E$ is the dimensionless deflection strength, with $V$ the deflecting voltage, $\lambda$ the RF field wavelength and $E$ the central beam energy. $R_{12}$ is the angular-to-spatial element of the transfer matrix from the deflector to the screen, for simplicity, $R_{12}$ is represented as $L_d$, the free drift length in the following discussions. It's straightforward to calculate the transverse beam size on the screen,

$$\sigma_x = \sqrt{\sigma_{xs}^2 + \sigma_z^2 u^2 L_d^2} \tag{2}$$

As is already known, the intrinsic beam size on the screen $\sigma_{xs}$ is a crucial factor to limit the resolution of the deflector. In order to eliminate the influence of transverse beam size, an additional TGU[23, 24] is utilized before the deflector to introduce a transverse-longitudinal coupling to the electron beam phase space[25, 26], i.e., *x-z* coupling here. As shown in Fig. 1, the *x* direction of the electron beam is transformed to *z*-related in a TGU. Then the transverse kick provided by the deflector and the following rotations in free space result in a *z*-dependent tilt of the beam. By properly setting all the parameters, the temporal profile of the electron beam is exactly mapped to the transverse plane without any reduction of the intrinsic beam size. This technique allows measuring the electron beam with a better temporal resolution. In other word, the required deflection voltage and/or RF frequency may be greatly relaxed for a targeted resolution.

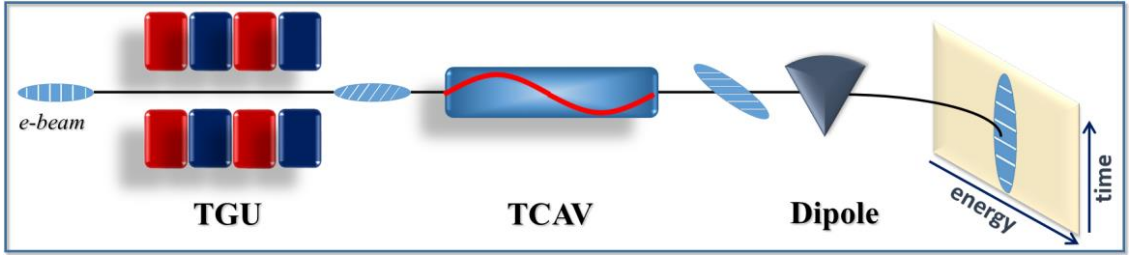

**Figure 1**. The schematic layout of the modified transverse deflector configuration. It includes the transverse gradient undulator, the transverse deflector, the magnetic spectrometer and the Ce:YAG screen located downstream of the FEL undulator. The deflecting structures provide horizontal streaking followed by a vertically bending dipole for measuring the energy spectrum. A camera captures the transverse beam profile on the diagnostic screen.

Considering a TGU with vertical magnetic field, total length of $L_u$, transverse gradient of α, and dimensionless undulator parameter of K, the transport matrix of the TGU can be approximately written as:

$$R_{TGU} = \begin{vmatrix} 1 & L_u & 0 & 0 & 0 & -L_u\tau/2 \\ 0 & 1 & 0 & 0 & 0 & -\tau \\ 0 & 0 & 1 & L_u & 0 & 0 \\ 0 & 0 & 0 & 1 & 0 & 0 \\ \tau & L_u\tau/2 & 0 & 0 & 1 & 0 \\ 0 & 0 & 0 & 0 & 0 & 1 \end{vmatrix},  \quad (3)$$

where $\tau = L_u K^2 \alpha / 2\gamma^2$ is the normalized gradient parameter of TGU. According to the single-particle dynamics, the horizontal beam position on the YAG screen can be described as

$$x = x_0(1 + u\tau L_d) + x_0'(L_d + L_u + u\tau L_d L_u/2) + z_0 u L_d - \tau\delta_0(L_d + L_u/2). \quad (4)$$

It turns out that when the parameters of TGU and the deflecting voltage satisfy $\tau = -1/uL_d$, the particle's position at the diagnostic screen is independent of the initial transverse position, and only depends on its initial longitudinal position, the initial divergence, and the initial energy deviation. Following the above derivation, the transverse beam profile on the screen is written as,

$$\sigma_x = \sqrt{(L_d + L_u/2)(\tau^2\sigma_\gamma^2 + \sigma_{x'}^2) + \sigma_z^2 u^2 L_d^2}. \quad (5)$$

In comparison with Eq. (2) and (5) for the typical beam parameters of modern high-gain FELs, the temporal resolution limitations from the beam divergence and beam energy spread is at approximately 1~2 order of magnitude less than that from the beam size. It is clearly obtained that the fundamental temporal resolution limit arising from the transverse beam size in the bunch profile measurement is significantly relaxed by the TGU enhanced deflector configuration.

**FEL temporal profile measurement.** Dalian Coherent Light Source (DCLS) is designed as an FEL user facility with continuously tunable wavelength of 50-150 nm[30], on the basis of high-gain harmonic generation (HGHG)[31], optical parametrical amplification laser and variable gap undulators. What's more, tapered undulator[32, 33] technique will also be utilized to further enhance the pulse energy to hundreds of micro-joule level. By the combination of rapid tuning and ultra-high intensity, DCLS is expected to enable the study of chemical reaction dynamics, probe matter at extremely small materials and surfaces, nanostructures and physical science. Inspired by the experiments at LCLS[20], DCLS is now considering FEL pulse profile characterization with a 1m X-band deflector located at the exit of undulators. And to achieve a better resolution, a variable gap TGU with 10cm period and 1m length is also suggested before the deflector.

The electron beam energy is 300 MeV with the normalized emittance of 1.0μm-rad, sliced energy spread of 30keV, bunch charge of 500pC, and peak current of 300A at DCLS. A representative HGHG setup is fully coherent 100nm FEL generation from a 300nm seed pulse of 100fs (FWHM) duration. In order to demonstrate the procedure of measurement and temporal reconstruction, start-to-end numerical tracking (see in Methods) including all the components of DCLS has been carried out, and we use an example in which the electron beam is over-bunched by strong energy modulation and a spiky FEL temporal profile[34] is generated in 6 segments of radiator undulator with segment length of 3m and period length of 30mm. For an HGHG scheme, the FEL lasing process can be easily suppressed by perturbing the radiator resonance. Then the 'FEL-off' longitudinal phase space image on the screen and the 'FEL-on' image for normal operation can be recorded. Fig. 2(a) and (b) shows the region of interest from the "FEL-off" and "FEL-on" image, respectively, with the deflection voltage of 12MV, the drift length of 1.5m, and the optimized TGU gradient of 50m$^{-1}$ and undulator parameter of K=20. The time-resolved energy loss and energy spread growth due to the FEL lasing process can be clearly obtained.

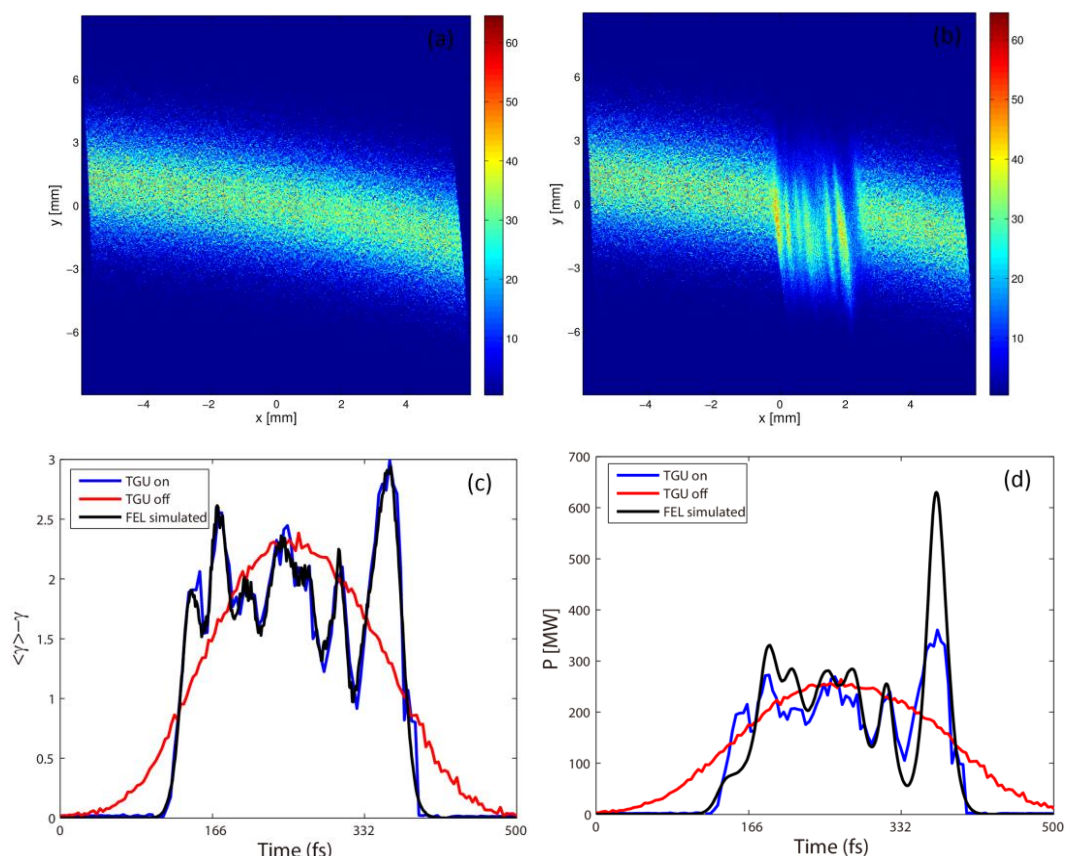

**Figure. 2.** The single-shot beam longitudinal phase space images (region of interest) are shown in (a) "FEL off" and (b) "FEL on". The beam energy loss from FEL simulation and from the deflector reconstructions w/o TGU is shown in (c). The FEL radiation profile from FEL simulation and from the deflector reconstructions w/o TGU is shown in (d).

To reconstruct the FEL temporal profile, the beam images are firstly sliced along the time dimension to get the time-dependent centroid energy $E(t)$ and current $I(t)$ in each time slice. While the horizontal projection of the beam images onto time represents the electron bunch current profile $I(t)$, the time-resolved mean energy loss $\Delta E(t)$ can be obtained by comparing the lasing-on and lasing-off sliced data. Then the absolute power profile can be retrieved directly as $P(t)= \Delta E(t) \times I(t)/e$. Fig. 2(c) shows the centroid energy loss $\Delta E(t)$, from which one may clearly see that the energy loss diagnostic with temporal resolution down to a few femtosecond is observed by the TGU enhanced deflector technique. The reconstructed FEL pulses are represented in Fig. 2(d) and compared with the true profile from FEL simulations. Considering that the e-beam transverse distribution and slippage of the radiation pulse relative to the electron beam are neglected in the reconstruction here, which actually is already some fraction of the achieved temporal resolution for a EUV wavelength FEL, a reasonable consistent FEL profile can be obtained by the TGU enhanced deflector technique. It is worth stressing that the FEL-on images can be subsequently recorded during normal FEL operation, thus permitting continuous, single-shot, time-resolved FEL temporal profile measurements.

**FEL micro-bunching monitor.** With this advanced deflector technique proposed, the accelerator performance and FEL lasing process can be studied and improved more effectively. For example, in comparison with the poor temporal coherence from self-amplified spontaneous emission, various external laser seeded schemes[25, 31, 35] have been proposed and developed for fully coherent FEL pulses generation. The first step of these seeded FELs is always the interaction between the seed laser and electron beam in short modulator, in which a 'footprint' left by the external seed laser will be included in the electron beam longitudinal phase space. A followed dispersive chicane then transfers the 'footprint' into a density modulation (the so-called 'micro-bunching'). Finally, micro-bunching excites coherent emission and even greater density modulation in the radiator. As a result, it leads to a net energy transfer from the electrons to coherent photons. Therefore, one of the key issues for seeded FELs is to achieve three-dimensional overlap and undulator resonance of the electron beam and the seed laser pulse in the modulator, and obtain a significant micro-bunching. In most of the early experiments[36-39], the laser-induced micro-bunching is confirmed by coherent radiation from the electron beam. Here we expect to monitor the micro-bunching itself by the deflector diagnostic scheme with high temporal resolution.

As an example, we discuss the laser-induced micro-bunching measurement for DCLS. Using the 300nm seed laser as a known ruler, the temporal resolution for accurate measurement of the micro-bunching in phase space should be less than 1 fs. According to Eq. (5), the TGU enabled deflector technique is immune to the transverse beam size completely, while still affected by the beam transverse divergence. One may add a telescope beam line before TGU to increase the ratio of the beam size to beam divergence and, hence, to reduce the influence of the beam divergence. In the following study, for simplicity, a 100μm electron beam size and an emittance of 0.2μm-rad which may be optimistically achieved in 50pC low charge mode of DCLS is considered.

Fig. 3 (a) shows the micro-bunching image calculated on the screen 5.0m downstream of the deflector, with the deflection voltage of 30MV, the optimized TGU gradient of 50m$^{-1}$ and undulator parameter of K=10. The fine structure of the electron beam density modulation is clearly revealed. The Fourier transform of the measured beam density distribution yield the bunching factor at the harmonics of the seed laser wavelength, as illustrated in Fig. 3 (b). The difference between the true micro-bunching from FEL simulation and that reconstructed from the deflector is caused by the limited resolution. The observation of the 3$^{rd}$ harmonic suggests that the temporal resolution is far less than 1fs, which enable the study of more subtle coherent phase space features. In the absence of TGU case, the micro-bunching is indistinguishable as shown in Fig. 3 (c), and an equivalent RF voltage over 100 MV is required to achieve a similar resolution as in Fig. 3 (a). In addition, the temporal resolution expected from Eq. (2) and (5) for the situation here is shown in Fig. 3(d), which agrees well with the above simulation results.

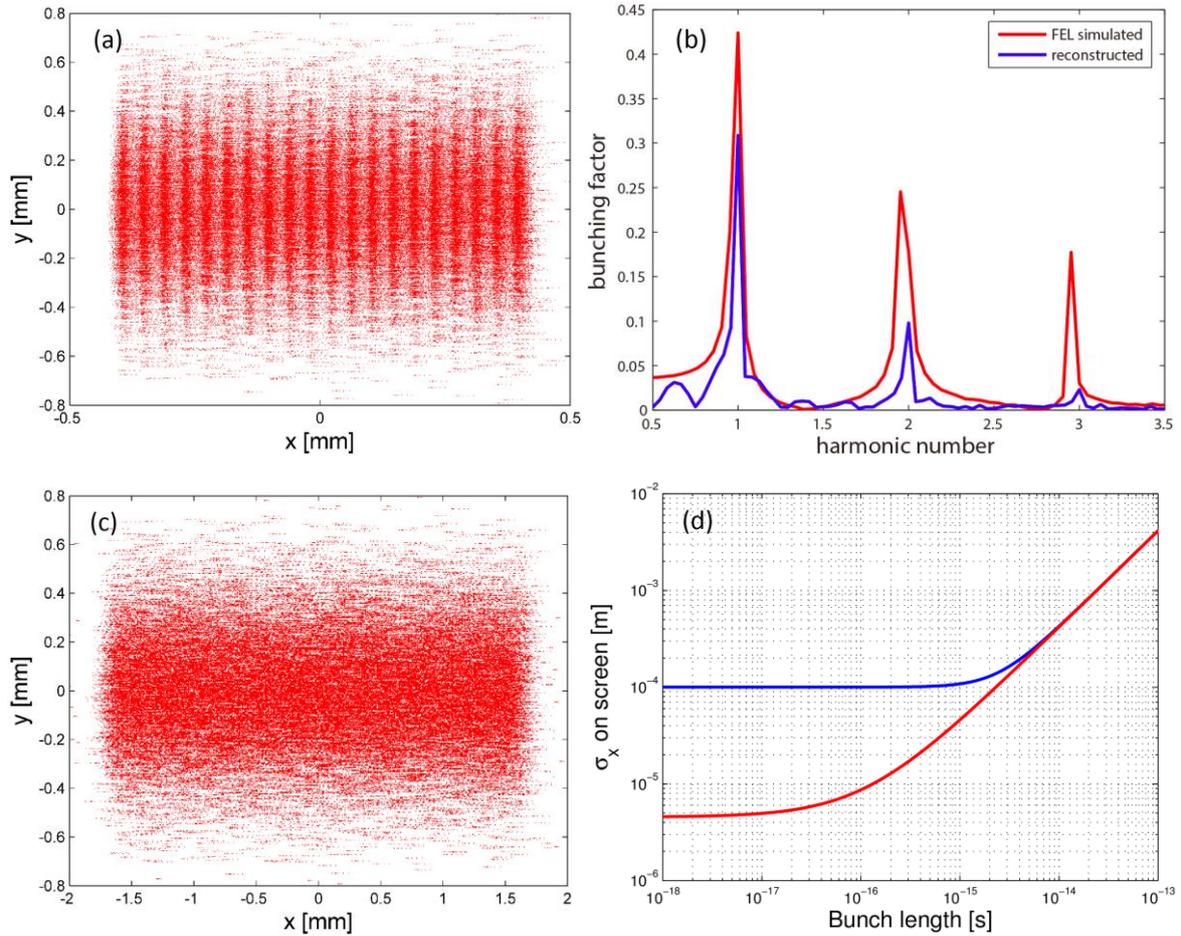

**Figure 3.** The calculated image on YAG screen downstream the deflecting cavity and the converted electron beam micro-bunching is shown in (a) and (b). For comparison study, a case without TGU is given in (c). Theoretical temporal resolution performances of two deflector configurations for this example is plotted in (d), where the blue curve is from Eq. (2) dots and the red one is from Eq. (5).

## Discussions

In summary, we have proposed a novel technique of the transverse deflector, in which the temporal resolution is enhanced by a transverse gradient undulator. It is capable of measuring the longitudinal beam profile with the resolution down to sub-femtosecond. As far as our knowledge, the function of TGU in this proposal is more or less same as a dipole magnet, thus TGU can be replaced by a dipole in principle. The advantage of TGU is that it allows a straight beam line and could avoid many constrains in the installation of the different components. In addition, TGU may also serve as a THz afterburner for pump-probe FEL experiment and other specified beam diagnostics.

It's also benefit to use the high resolution transverse deflector for optimize the energy modulation process in seeded FELs, which will provide an efficient tool for the study of the FEL lasing process and improvement of the machine performance. Finally, it is worth emphasizing that the temporal resolution of the proposed technique is still related to the beam divergence, even the effects of transverse beam size is already eliminated. Thus for real accelerators, the beam size can be artificially increased to further improve the performance.

## Methods

**Fabrication and measurement of a TGU prototype.** TGU has been proposed to reduce the sensitivity to electron energy jitters for FEL oscillators[23] and improve the performances for high-gain FEL amplifiers[24-26]. Typically, the transverse gradient of magnetic field may be obtained either by tilting the two magnetic poles relative in Halbach type hybrid undulators or by using the curvature of a cylindrically shaped undulator body, e.g., in superconducting coils[40]. More recently, we have constructed a variable gap planar undulator with 60mm period length and 0.60m total length, in which magnetic poles with 2mm height step and 45 degree slope are utilized to obtain the designed transverse gradient. According to the measurement, the magnetic peak field is about 1.2T, the achieved transverse gradient is more than 50T/m at an 8 mm magnetic gap, and the beam trajectory straightness in this TGU prototype due to transverse gradient can be successfully compensated by a 20Gs uniform magnetic field produced by a long coil.

**Numerical simulations.** Collective effects including those from accelerating structures, undulators, transverse deflector, and coherent synchrotron radiation in the spectrometer dipoles are considered in the tracking process. The electron beam dynamics in photo-injector is simulated with ASTRA[41] to take into account the space-charge effects and ELEGANT[42] is used for the simulation in the remainder of the linac. The universal FEL code GENESIS[43] is utilized to calculate FEL performances and micro-bunching process, then the beam dynamics in TGU is numerically tracked with a 6-D transfer matrix. At last, the electron beam is imported into ELEGANT again to track the particle in the deflector and spectrometers.

## Acknowledgement

The authors are grateful to Chao Feng, Bo Liu and Zhirong Huang for helpful discussions. This work was supported by the National Natural Science Foundation of China (21127902, 11175240, 11205234 and 11322550).

## Author Contributions

H. D. conceived the novel design of deflector. G.W. and M.Z. performed the numerical simulations. W.Z. fabricated the TGU prototype and performed the measurements. G.W., H.D. and X.Y. co-wrote the paper with inputs from all authors.

## Additional information

**Competing financial interests:** The authors declare no competing financial interests.